\documentclass[12pt]{article}
\begin{document}
\thispagestyle{empty}
\centerline{\Large Stability of matter in the accelerating spacetime}

\bigskip\bigskip
\centerline{Miroslaw Kozlowski$^{\rm a,c}$, Janina Marciak-Kozlowska$^{\rm
b}$}

\bigskip
\noindent{\llap{$^{\rm a}$~} Institute of Experimental Physics and Science Teacher College of Warsaw University,  Hoza~69, 00-681 Warsaw, Poland, e-mail: mirkoz@ids.pl}

\noindent{\llap{$^{\rm b}$~}Institute of Electron Technology,
   Al.~Lotnikow~32/46, 02-668 Warsaw, Poland}

\hbox to 5cm{\hsize=5cm\vbox{\ \hrule}}\par
\noindent{\llap{$^{\rm c}$~}Author to whom correspondence should be addressed.}

\newpage
\section{Introduction}
In the seminal paper~\cite{1} F.~Calogero described the cosmic origin of
quantization. In paper~\cite{1} the tremor of the cosmic particles is the
origin of the quantization and the characteristic acceleration of these
particles $a\sim10^{-10}$~m/s$^2$ was calculated. In our earlier
paper~\cite{2} the same value of the acceleration was obtained and
compared to the experimental value of the measured spacetime
acceleration. In this paper we define the cosmic force --\textit{Planck}
force, $F_{Planck}=M_Pa_{Planck} \;(a_{Planck}\sim a)$ and study the
history of Planck force as the function of the age of the Universe.

Masses introduce a curvature in spacetime, light and matter are forced to
move according to spacetime metric. Since all the matter is in motion,
the geometry of space is constantly changing. A Einstein relates the
curvature of space to the mass/energy density:
    \begin{equation}
    \mathbf{G}=k\mathbf{T}.\label{eq1}
    \end{equation}
$\mathbf{G}$ is the Einstein curvature tensor and $\mathbf{T}$ the
stress-energy tensor. The proportionality factor $k$ follows by
comparison with Newton's theory of gravity: $k=G/c^4$ where $G$ is the
Newton's gravity constant and $c$ is the vacuum velocity of light; it
amounts to about $2.10^{-43}\;{\rm N}^{-1}$, expressing the {\it
rigidity} of spacetime.

In paper~\cite{2} the model for the acceleration of spacetime was
developed. Prescribing the $-G$ for spacetime and $+G$ for matter the
acceleration of spacetime was obtained:
\begin{equation}
a_{
Planck}=-\frac12\left(\frac{\pi}{4}\right)^{1/2}\frac{(N+\frac34)^{1/2}}
{M^{3/2}}A_P,\label{eq2}
\end{equation}
where $A_P$, \textit{Planck} acceleration equal, viz.:
\begin{equation}
A_P=\left(\frac{c^7}{\hbar
G}\right)^{1/2}=\frac{c}{\tau_P}\cong10^{51}ms^{-2}.\label{eq3}
\end{equation}
As was shown in paper~\cite{2} the $a_{Planck}$ for $N=M=10^{60}$ is of
the order of the acceleration detected by Pioneer spacecrafts~\cite{3}.

Considering $A_P$ it is quite natural to define the \textit{Planck} force
$F_{Planck}$,
\begin{equation}
F_{Planck}=M_PA_P=\frac{c^4}{G}=k^{-1},\label{eq4}
\end{equation}
where
$$
M_P=\left(\frac{\hbar c}{G}\right)^{1/2}.
$$
From formula~(\ref{eq4}) we conclude that $F^{-1}_{Planck}=$ rigidity of
the spacetime. The \textit{Planck} force,
$F_{Planck}=c^4/G=1.2\;10^{44}$~N can be written in units which
characterize the microspacetime, i.e. GeV and fm.

In that  units
$$
k^{-1}=F_{Planck}=7.6\;10^{38}\; {\rm GeV/fm}.
$$

\section{The \textit{Planck}, \textit{Yukawa} and \textit{Bohr} forces}
As was shown in paper~\cite{2} the present value of \textit{Planck} force
equal
    \begin{equation}
    F^{Now}_{Planck}(N=M=10^{60})\cong-\frac12\left(\frac{\pi}{4}\right)^{1/2}
    10^{-60}\frac{c^4}{G}=-10^{-22}{\rm \frac{GeV}{fm}}.\label{eq5}
    \end{equation}
In paper~\cite{4} the \textit{Planck} time $\tau_P$ was defined as the
relaxation time for spacetime
    \begin{equation}
    \tau_P=\frac{\hbar}{M_Pc^2}.\label{eq6}
    \end{equation}
Considering formulae~(\ref{eq4}) and (\ref{eq6}) $F_{Planck}$ can be
written as
    \begin{equation}
    F_{Planck}=\frac{M_Pc}{\tau_P},\label{eq7}
    \end{equation}
where $c$ is the velocity for gravitation propagation. In paper~\cite{5}
the velocities and relaxation times for thermal energy propagation in
atomic and nuclear matter were calculated:
    \begin{eqnarray}
    v_{atomic}&=&\alpha_{em}c,\label{eq8}\\
    v_{nuclear}&=&\alpha_sc,\nonumber
    \end{eqnarray}
where $\alpha_{em}=e^2/(\hbar c)=1/137, \; \alpha_s=0.15$. In the
subsequent we define atomic and nuclear accelerations:
    \begin{eqnarray}
    a_{atomic}&=&\frac{\alpha_{em}c}{\tau_{atomic}},\label{eq9}\\
    a_{nuclear}&=&\frac{\alpha_s c}{\tau_{nuclear}}.\nonumber
    \end{eqnarray}
Considering that $\tau_{atomic}=\hbar/(m_e\alpha^2_{em}c^2)$,
$\tau_{nuclear}=\hbar/(m_N\alpha^2_sc^2)$ one obtains from
formula~(\ref{eq9})
    \begin{eqnarray}
    a_{atomic}&=&\frac{m_ec^3\alpha^3_{em}}{\hbar},\label{eq10}\\
    a_{nuclear}&=&\frac{m_Nc^3\alpha^3_s}{\hbar}.\nonumber
    \end{eqnarray}
We define, analogously to \textit{Planck} force the new forces:
$F_{Bohr}$, $F_{Yukawa}$
    \begin{eqnarray}
    F_{Bohr}&=&m_ea_{atomic}=\frac{(m_ec^2)^2}{\hbar c}\alpha^3_{em}=
    5\;10^{-13}\;{\rm \frac{GeV}{fm}},\label{eq11}\\
    F_{Yukawa}&=&m_Na_{nuclear}=\frac{(m_Nc^2)^2}{\hbar c}\alpha^3_s=
    1.6\;10^{-2}\;{\rm \frac{GeV}{fm}}.\nonumber
    \end{eqnarray}
Comparing formulae~(\ref{eq5}) and (\ref{eq11}) we conclude that
gradients of \textit{Bohr} and \textit{Yukawa} forces are much large than
$F^{Now}_{Planck}$, i.e.:
    \begin{eqnarray}
    \frac{F_{Bohr}}{F^{Now}_{Planck}}&=&\frac{5.10^{-13}}{10^{-22}}\sim10^9,
    \label{eq12}\\
    \frac{F_{Yukawa}}{F^{Now}_{Planck}}&=&\frac{10^{-2}}{10^{-22}}\sim10^{20}.
    \nonumber
    \end{eqnarray}
The formulae~(\ref{eq12}) guarantee present day stability of matter on
the nuclear and atomic levels.

As the time dependence of $F_{Bohr}$ and $F_{Yukawa}$ are not well
established, in the subsequent we will assumed that $\alpha_s$ and
$\alpha_{em}$ do not dependent on time. Considering formulae~(\ref{eq8})
and (\ref{eq11}) we obtain
    \begin{eqnarray}
    \frac{F_{Yukawa}}{F_{Planck}}&=&\frac{1}{(\frac{\pi}{4})^{1/2}}
    \frac{(m_Nc^2)^2}{M_Pc^2}\frac{\alpha^3_s}{\hbar}T,\label{eq13}\\
    \frac{F_{Bohr}}{F_{Planck}}&=&\frac{1}{(\frac{\pi}{4})^{1/2}}
    \frac{(m_ec^2)^2}{M_Pc^2}\frac{\alpha_{em}}{\hbar}T.\label{eq14}
    \end{eqnarray}
As can be
realized from formulae~(\ref{eq13}), (\ref{eq14}) in the past $F_{Planck}\sim F_{Yukawa}$ (for
$T=0.002$~s) and $F_{Planck}\sim F_{Bohr}$ (for $T\sim 10^8$~s), $T=$ age
of universe. The calculated ages define the limits for instability of the
nuclei and atoms.

\section{The \textit{Planck}, \textit{Yukawa} and \textit{Bohr} particles}
In 1900 M. Planck~\cite{6} introduced the notion of the universal mass,
later on called the \textit{Planck} mass
    \begin{equation}
    M_P=\left(\frac{\hbar c}{G}\right)^{1/2},\qquad F_{Planck}=
    \frac{M_Pc}{\tau_P}.\label{eq15}
    \end{equation}
Considering the definition of the \textit{Yukawa} force~(\ref{eq11})
    \begin{equation}
    F_{Yukawa}=\frac{m_Nv_N}{\tau_N}=\frac{m_N\alpha_{strong} c}
    {\tau_N},\label{eq16}
    \end{equation}
the formula~(\ref{eq16}) can be written as:
    \begin{equation}
    F_{Yukawa}=\frac{m_{Yukawa} c}{\tau_N},\label{eq17}
   \end{equation}
where
    \begin{equation}
    m_{Yukawa}=m_N\alpha_{strong}\cong 147\;
{\rm \frac{MeV}{c^2}}\sim m_{\pi}.
    \label{eq18}
    \end{equation}
From the definition of the \textit{Yukawa} force we deduced the mass of
the particle which mediates the strong interaction -- pion mass
postulated by Yukawa in~\cite{7}.

Accordingly for \textit{Bohr} force:
    \begin{eqnarray}
    F_{Bohr}&=&\frac{m_e v}{\tau_{Bohr}}=\frac{m_e\alpha_{em}c}{\tau_{Bohr}}=
    \frac{m_{Bohr} c}{\tau_{Bohr}},\label{eq19}\\
    m_{Bohr}&=&m_e\alpha_{em}=3.7 \;{\rm \frac{keV}{c^2}}.\label{eq20}
    \end{eqnarray}
For the \textit{Bohr} particle the range of interaction is
    \begin{equation}
   r_{Bohr}=\frac{\hbar}{m_{Bohr}c}\sim0.1 \;{\rm nm},\label{eq21}
    \end{equation}
which is of the order of atomic radius.

Considering the electromagnetic origin of the mass of the \textit{Bohr}
particle, the planned sources of hard electromagnetic field, e.g. free
electron laser (FEL)at TESLA accelerator (DESY)~\cite{8} are best suited
to the investigation of the properties of the \textit{Bohr} particles.

\section{Possible interpretation of $F_{Planck}$, $F_{Yukawa}$ and $F_{Bohr}$.}
In an important work, published already in 1951 J.~Schwinger~\cite{9}
demonstrated that in the background of a static uniform electric field,
the QED vacuum is unstable and decayed with spontaneous emission of
$e^+e^-$ pairs. In the paper~\cite{9} Schwinger calculated the critical
field strengths $E_S$:
    \begin{equation}
    E_S=\frac{m_e^2c^3}{e\hbar}.\label{eq22}
    \end{equation}
Considering formula~(\ref{eq22}) we define the  \textit{Schwinger}
force:
    \begin{equation}
    F^e_{Schwinger}=eE_S=\frac{m^2_ec^3}{\hbar}.\label{eq23}
    \end{equation}
Formula~(\ref{eq23}) can be written as:
    \begin{equation}
    F^e_{Schwinger}=\frac{m_ec}{\tau_{Sch}},\label{eq24}
    \end{equation}
where
    \begin{equation}
    \tau_{Sch}=\frac{\hbar}{m_ec^2}\label{eq25}
    \end{equation}
is \textit{Schwinger}  relaxation time for the creation of $e^+e^-$ pair.
Considering formulae~(\ref{eq11}) the relation of $F_{Yukawa}$ and
$F_{Bohr}$ to the \textit{Schwinger}  force can be established
    \begin{eqnarray}
    F_{Yukawa}&=&\alpha^3_s\left(\frac{m_N}{m_e}\right)^2
    F^e_{Schwinger},\qquad\alpha_s=
    0.15,\label{eq26}\\
    F_{Bohr}&=&\alpha^3_{em}F^e_{Schwinger},\qquad \alpha_{em}=\frac{1}{137},\nonumber
    \end{eqnarray}
and for \textit{Planck} force
    \begin{equation}
    F_{Planck}=\left(\frac{M_P}{m_e}\right)^2F^e_{Schwinger}.\label{eq27}
    \end{equation}

In Table~1 the values of the $F^e_{Schwinger}$, $F_{Planck}$,
$F_{Yukawa}$ and $F_{Bohr}$ are presented, all in the same units GeV/fm.
As in those units the forces span the range $10^{-13}$ to $10^{38}$ it is
valuable to recalculate the \textit{Yukawa} and \textit{Bohr} forces in
the units natural to nuclear and atomic level. In that case one obtains:
    \begin{equation}
    F_{Yukawa}\sim\frac{16\;{\rm MeV}}{\rm fm}.\label{eq28}
    \end{equation}
It is quite interesting that $a_v\sim16$~MeV is the volume part of the
binding energy of the nuclei (droplet model).

\begin{table}[h]
\caption{Schwinger, \textit{Planck}, \textit{Yukawa} and \textit{Bohr}
forces}
\begin{center}
\begin{tabular}{|c|c|c|c|}
\hline
 $F_{Schwinger}^e$&$F_{Planck}$~&$F_{Yukawa}$~&
$F_{Bohr}$~\\
\hline
[GeV/fm]&[GeV/fm]&[GeV/fm]&[GeV/fm]\\
\hline $\sim10^{-6}$&$\sim10^{38}$
&$\sim10^{-2}$&$\sim10^{-13}$\\
\hline
\end{tabular}
\end{center}
\end{table}

For the \textit{Bohr} force considering formula~(\ref{eq11}) one obtains:
    \begin{equation}
    F_{Bohr}\sim\frac{50\;{\rm eV}}{0.1\;{\rm nm}}.\label{eq29}
    \end{equation}
Considering that the Rydberg energy $\sim 27$~eV and \textit{Bohr} radius
$\sim0.1$~nm formula~(\ref{eq29}) can be written as
    \begin{equation}
    F_{Bohr}\sim\frac{\rm Rydberg \; energy}{\rm Bohr \; radius}\label{eq30}
\end{equation}
\section{Concluding remarks}
In this paper the forces: \textit{Planck}, \textit{Yukawa} and
\textit{Bohr} were defined. It was shown that the present value of the
\textit{Planck} force (which is the source of the universe acceleration)
$\sim10^{-22}$~GeV/fm is much smaller than the \textit{Yukawa}
($\sim10^{-2}$~GeV/fm) and \textit{Bohr} ($10^{-13}$~GeV/fm) forces
respectively. This fact guarantees the stability of the matter in the
present. However in the past for $T$ (age of the universe), $T<0.002$~s,
$F_{Yukawa}<F_{Planck}$ (0.002 s) and $F_{Bohr}<F_{Planck}$ ($10^8$ s).
In this paper the relation of the \textit{Schwinger}  force (for the
vacuum creation of the $e^+e^-$ pairs) to the \textit{Planck},
\textit{Yukawa} and \textit{Bohr} force was obtained.

\newpage

\end{document}